\documentclass[letterpaper,10pt]{revtex4}
\usepackage{graphicx}
\begin{document}

\title{Measuring the Earth's gravity field with cold atom interferometers}
\author{Olivier Carraz}
\author{Christian Siemes}
\author{Luca Massotti}
\author{Roger Haagmans}
\author{Pierluigi Silvestrin}
\affiliation{Earth Observation Programmes - European Space Agency \\
ESTEC P.O. Box 299, 2200 AG Noordwijk, The Netherlands \\
E-mail: olivier.carraz@esa.int - Fax: +31 (0) 71 565 4696}

\keywords{Atom interferometry; Gravity Gradiometry; Gyroscope; Next Generation Satellite Gravimetry Mission}

\begin{abstract}
The scope of the paper is to propose different concepts for future space gravity missions using Cold Atom Interferometers (CAI) for measuring the diagonal elements of the gravity gradient tensor, the spacecraft angular velocity and the spacecraft acceleration. The aim is to achieve better performance than previous space gravity missions due to a very low white noise spectral behaviour of the CAI instrument and a very high common mode rejection, with the ultimate goals of determining the fine structures of the gravity field with higher accuracy than GOCE and detecting time-variable signals in the gravity field.
\end{abstract}

\maketitle

\section{Introduction}
In the past decades, it has been shown that atomic quantum sensors have the potential to drastically increase the performance of inertial measurements \cite{{Peters},{Sorrentino_gradio}}. These inertial sensors present a very low and spectrally white noise as opposed to classical accelerometers, which present colored noise. Recent results in different labs show that it is possible to build a reliable system for using atom interferometry on vehicles and for space applications. These last developments prove that this technology, which is already suitable on ground vehicles \cite{{Wu},{Bidel}}, is competitive with classic inertial sensors. Some developments already worked in zero-g environment in the drop tower facility in Bremen, Germany \cite{Muntinga}, or in a 0-g plane \cite{Geiger} and are the state-of-the-art of compact setups aiming for spaceborne platform.

Particularly, we propose here a concept using CAI for measuring all diagonal elements of the gravity gradient tensor and the full spacecraft angular velocity vector in order to achieve better performance than the GOCE gradiometer over a larger measurement bandwidth \cite{Carraz}, with the ultimate goals of determining the fine structures and time-variable signals in the gravity field better than today. This concept relies on a high common mode rejection and a longer interaction time due to the micro gravity environment, which will provide a better performance than any other cold atom interferometer on the ground.

This CAI gravity gradiometer (GG) allows reaching a sensitivity of $4.7\;mE/Hz^{1/2}$, with the promise of a flat noise power spectral density also at low frequency, and a very high accuracy on rotation rates (below $35\;prad/s/Hz^{1/2}$). Thanks to high common mode rejection, drag free constraints will be relaxed with respect to the GOCE mission and non-gravitational forces will be measured by combining the different measurements from the cold atom interferometers.

Estimation of an Earth gravity field model from the new gravity gradiometer concept has to be evaluated taking into account different system parameters such as attitude control, altitude of the satellite, mission lifetime, etc.

Other use of CAI can be achieved by hybridizing quantum and electrostatic sensors. Such a hybridization could improve the sensitivity of inertial sensors used for Next Generation Gravity Missions (NGGM) \cite{{Silvestrin},{Massotti}} by benefiting from the performances of both technologies.

\section{Atom interferometry}
\begin{figure}
\centering \includegraphics[width=0.4\linewidth]{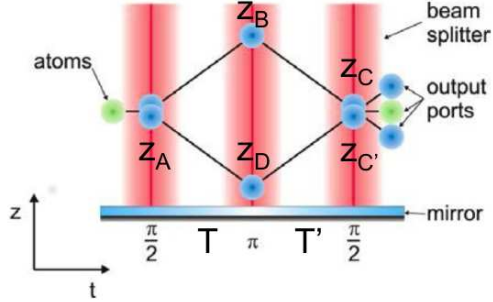}
   \caption{The double diffraction scheme for one cloud of atoms. Figure from \cite{Tino}. \label{Double_diffraction}}
 \end{figure}

Atom interferometers rely on the wave-particle duality, which allows matter waves to interfere, and on the superposition principle. They can be sensitive to inertial forces. This Chu-Bord\'e interferometer can be extended to any kind of inertial sensor such as a gravimeter \cite{Peters}, a gyroscope \cite{Canuel} or a gravity gradiometer \cite{Sorrentino_gradio}.
In a Chu-Bord\'e interferometer the test mass is a cloud of cold atoms, which is obtained from a Magneto-Optical Trap (MOT) by laser cooling and trapping techniques. This cloud of cold atoms is released from the trap and its acceleration due to external forces is measured by an atom interferometry technique. A Chu-Bord\'e interferometer consists in a sequence of three equally spaced Raman laser pulses \cite{Kasevich}, which drive stimulated Raman transitions between two stable states of the atoms. In the end, the proportion of atoms in the two stable states depends sinusoidally on the phase of the interferometer $\Phi$, which is proportional to the acceleration of the atoms along the Raman laser axis of propagation in the reference frame defined by the Raman mirror.
The Chu-Bord\'e interferometer in a double diffraction scheme (see Fig. \ref{Double_diffraction}) allows to enlarge the sensitivity by a factor 2, and to suppress at first order parasitic effects such as the light shift or the magnetic field, as the atoms remain in the same internal state \cite{{Leveque},{Giese}}.

\begin{figure}
\centering \includegraphics[width=0.8\linewidth]{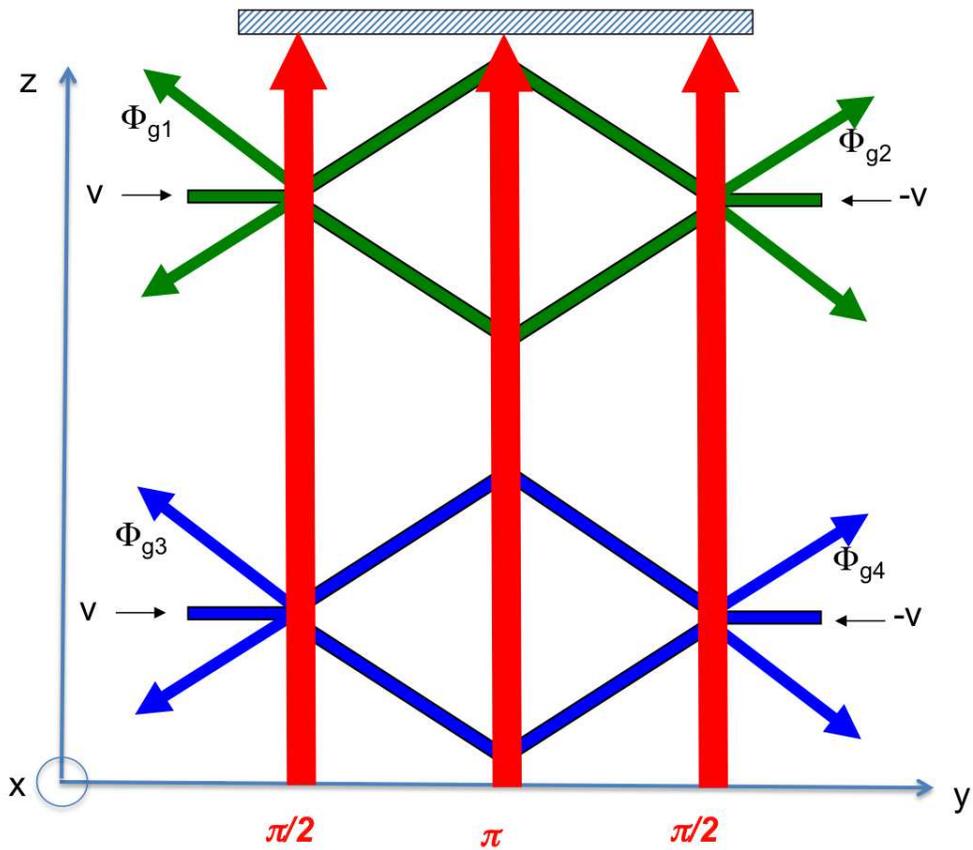}
   \caption{Combining 4 atom interferometers to measure both gravity gradient and rotation rate. \label{Concept_interfero}}
 \end{figure}

It is possible to suppress non-gravitational forces by measuring simultaneously different atom interferometers \cite{{Sorrentino_gradio},{Dickerson}}. On Fig. \ref{Concept_interfero} as the inertial reference being common to the 4 different atom interferometers, the high common rejection cancels the vibration noise and combining the phases $\Phi g_{1}$ and $\Phi g_{2}$ (or $\Phi g_{3}$ and $\Phi g_{4}$) we can precisely measure the rotation rate $\omega_{x}$, knowing precisely the velocity $\vec{v}$ of the atom cloud along y-axis. This is due to the measurement of the Coriolis effect $2\vec{\Omega}\times\vec{v}$ along z-axis. Combining  $\Phi g_{1}$ and $\Phi g_{3}$ (or $\Phi g_{2}$ and $\Phi g_{4}$) allows to measure $\gamma$, i.e. the gravity gradient combined with the centrifugal acceleration, relying on the precise knowledge of the distance between the atom clouds. Applying this measurements in the 3 orthogonal directions, the full angular velocity $\vec{\Omega}$ is simultaneously measured, thus we have direct access to all diagonal elements of the gravity gradient tensor $V_{xx},V_{yy},V_{zz}$.
 
\section{CAI GG Instrument concept}

  \begin{figure*}
\centering\includegraphics[width=0.8\linewidth]{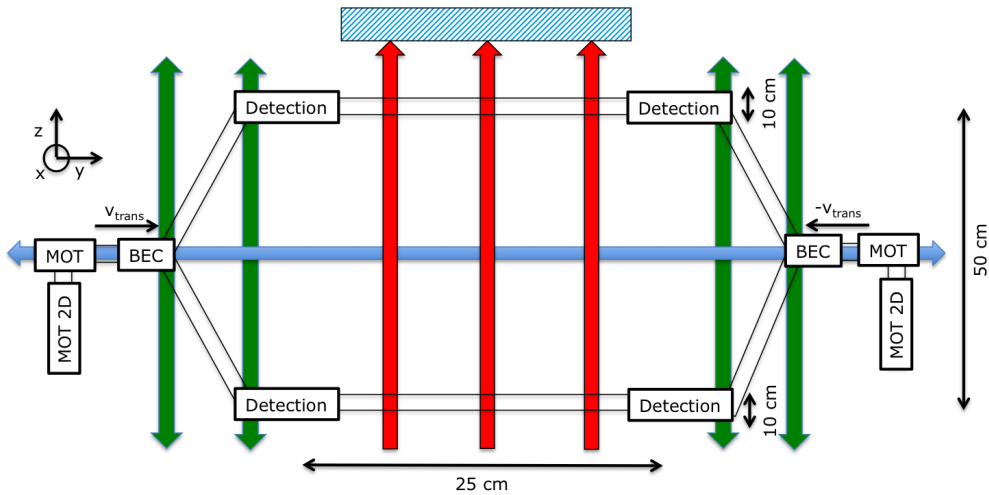}
   \caption{Scheme of the vacuum chamber (not drawn to scale; orientative sizing). The red arrows represent the Raman lasers for the interferometry part; the blue arrows represent the light pulse giving the transverse velocity $v_{trans}$; the green arrows represent the light pulses guiding the atoms from the cooling room to the interferometer room; the blue rectangle is the mirror, which will be the inertial reference.} 
 \label{Concept_system}  
 \end{figure*}
 
Fig. \ref{Concept_system} describes the gravity gradiometer concept in one dimension. This one-dimensional concept consists in measuring one diagonal element of the gravity tensor ($V_{zz}$) and the rotation rate along the x-axis. It can be extended to the other two dimensions in order to obtain the full diagonal elements of the gravity gradient tensor and the full angular rate vector. Full description of this instrument and a more detailed trade-off on the noise sources of the instrument have been done in \cite{Carraz}.
Considering that the interferometer noise is limited by the quantum projection noise with $N=10^{6}$ atoms (i.e. an interferometer phase noise proportional to $1/\sqrt{N}$ per shot), the performances expected for such an instrument is for $f<1/(2T)$::

\begin{equation}
\begin{array}{l l}
\Delta\gamma=4.7\;mE/\sqrt{Hz} \\
\Delta\omega=35\;prad.s^{-1}/\sqrt{Hz}
\end{array}
\end{equation}

\begin{figure}
\centering\includegraphics[width=0.8\linewidth]{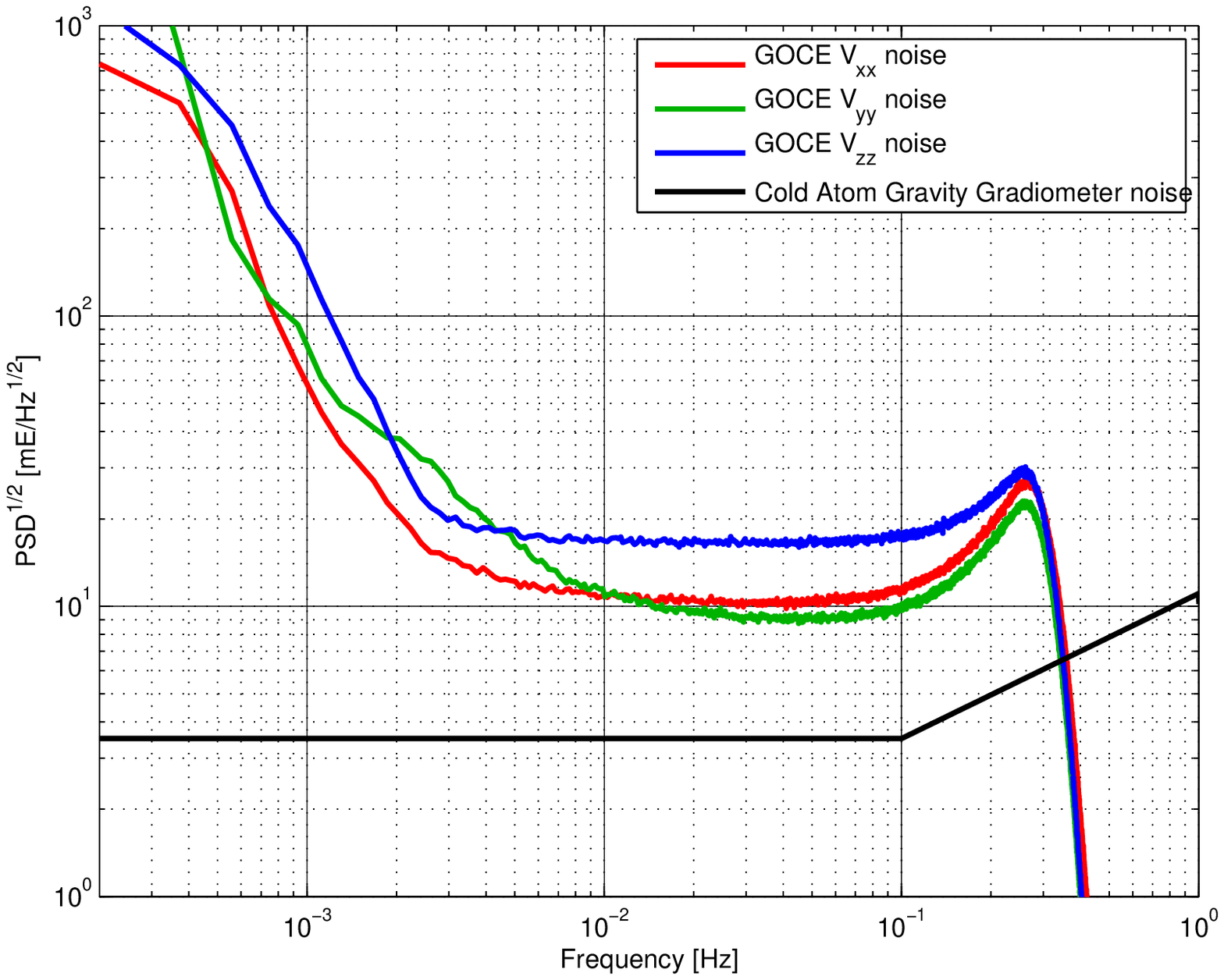}
\caption{Comparison between GOCE PSD and 1-D CAI GG PSD. \label{PSD}}
\end{figure}

As the time between each measurement ($T_{cycle}=1\;s$) is shorter than the interferometer time ($2T=10\;s$), for $1/(2T) < f < 1/T_{cycle}$ the sensitivity is increased by a factor $(2T.f)^{1/2}$ (see Fig. \ref{PSD}).

We present here the constraints on the control of the payload and the knowledge of the non-gravitational forces to obtain the performances of this CAI GG.

\subsection{Control of the payload}
We assume that to perform the best performances of the Cold Atom Interferometer (CAI) we need to control the position of the atoms better than $\delta z=1\;cm$ (extension of the cloud) at the end of the interferometer (detection zone). Thus the time of the propagation is:
\begin{equation}
\label{Tlong}
T_{prop}=T_{transport}+2T+T_{det}=12\;s
\end{equation} 

We then have:
\begin{equation}
\label{control1}
\delta a=\frac{2\delta z}{T_{prop}^{2}}= 1.4\times10^{-4}\;m.s^{-2}\\
\end{equation}

All other non-gravitational forces should imply lower acceleration than Eq. \ref{control1}, for instance for Coriolis force and angular acceleration we have:
\begin{equation}
\label{control2}
\begin{array}{l l}
\delta\Omega=\frac{\delta a}{2v_{trans}}= 2.7\times10^{-3}\;rad.s^{-1}\\
\delta\dot{\Omega}=\frac{\delta a}{R_{COM}}= 4.8\times10^{-4}\;rad.s^{-2}
\end{array}
\end{equation}
Where $R_{COM}$ is the distance between the atoms and the Raman mirror.

\subsection{Knowledge of the inertial forces}
As the measurement is the phase of an interferometer, in order to avoid any ambiguity there should not have any phase jump between two measurements ($T_{cycle}=1\;s$). This implies that each kind of measurements (e.g. vibration measurements $\Delta a$, rotation measurements $\Delta\Omega$, gravity gradient measurements $\Delta\Gamma$) has a limit scale between each interferometer sequence:
\begin{equation}
\label{know}
\begin{array}{l l}
\Delta a=\frac{2\pi}{k_{eff}.T^{2}}= 7.8\times10^{-9}\;m.s^{-2}\\
\Delta\Omega=\frac{2\pi}{2k_{eff}.v_{trans}.T^{2}}= 1.6\times10^{-7}\;rad.s^{-1}\\
\Delta\Gamma=\frac{\Delta a}{d}= 16\;E
\end{array}
\end{equation}

The knowledge of any inertial force should be better than these values within $T_{cycle}=1\;s$.

\section{First estimations of performance related to two mission scenarios}
Based on the instrument performances reported above, it has been proposed to retrieve the Earth geoid for two mission scenarios:
\begin{itemize}
\item The first scenario focuses on mapping the static gravity field at high accuracy and high spatial resolution by the means of high resolution gravity gradients. This implies that the orbital height is chosen to be very low at the cost of a short mission lifetime due to the drag forces to be compensated (Static gravity scenario);
\item The second scenario focuses on monitoring the time-variable gravity field, which ideally implies a mission lifetime around 10 years. Therefore, the orbital height is expected to be higher in this case in comparison to the static gravity scenario (Time-variable gravity scenario).
\end{itemize}

The simulations are based on The Updated ESA Earth System Model for Gravity Mission Simulation Studies (ESA ESM) \cite{Earthmodel} for the signal and an instrument noise of $4.7\;mE/\sqrt{Hz}$ for the CAI GG, a $2\;cm$ orbit accuracy and a $100\;nrad$ attitude control error due to the Satellite Track Ranging (STR) and CAI GG.
The background model errors considered are the $10\;\%$ AOHIS for the mean gravity field and $10\;\%$ AO for the time-variable gravity field.
The errors in tide models are approximated by 8 major tidal constituents of the difference between the two tide models EOT11a and GOT4.7.

\subsection{Static gravity scenario}

\begin{figure}
\centering \includegraphics[width=0.8\linewidth]{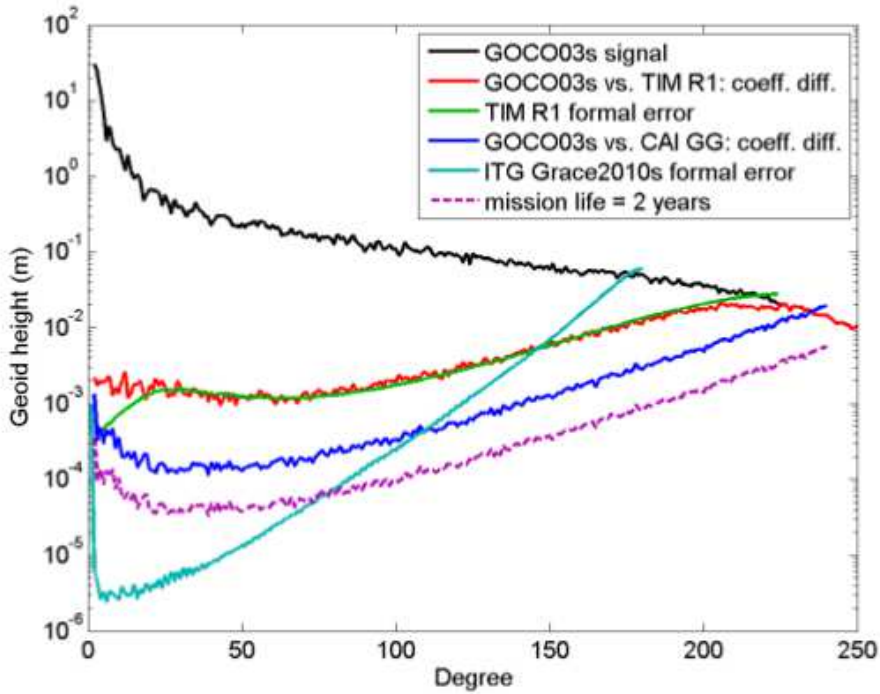}
\caption{Geoid height errors. \label{mean_gravity}}
\end{figure}

Considering a near polar circular orbit ($e=0.001$ and $i= 89^{o}$) at an altitude of 250 km, with a 61 day repeat, Fig. \ref{mean_gravity} shows that a 2-month solution allows an improvement by one order of magnitude with respect to GOCE mission. The accuracy is extrapolated for a 2-year mission which demonstrates a better accuracy of the geoid from order 80 to 250.

\subsection{Time-variable gravity scenario}

\begin{figure}
\centering \includegraphics[width=0.8\linewidth]{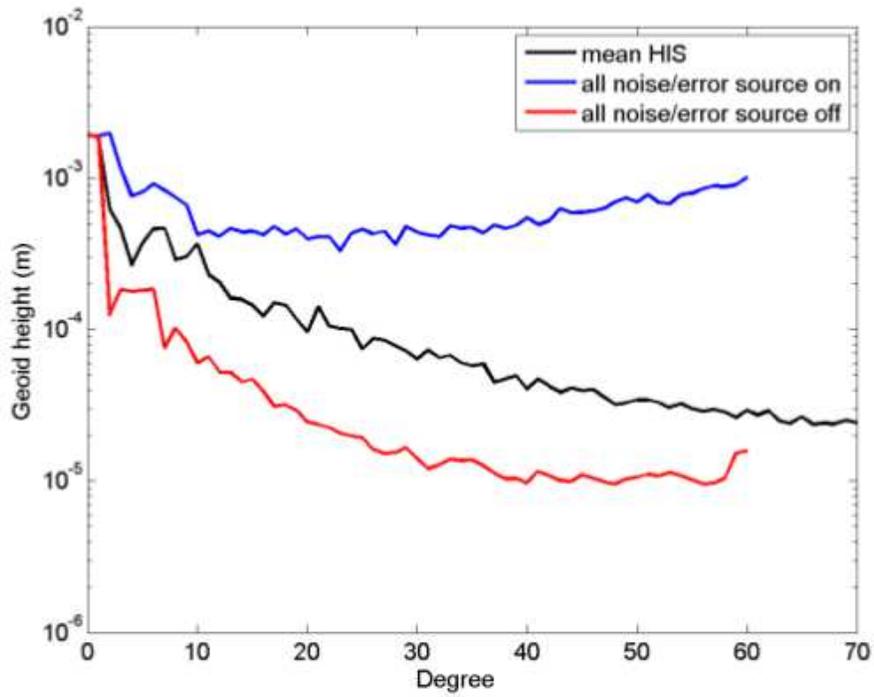}
\centering \includegraphics[width=0.8\linewidth]{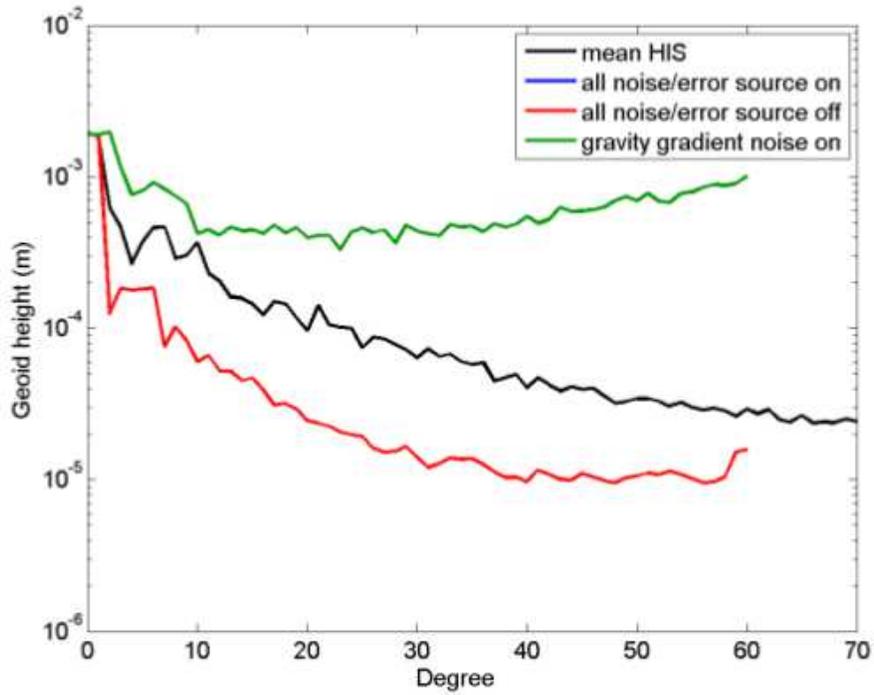}
\caption{Time variable geoid height error. Up: Comparison between with/without taking into account all errors. Down: Performance of the instrument with error coming only from the CAI GG.\label{time_gravity}}
\end{figure}

Considering a near polar circular orbit ($e=0.001$ and $i= 89^{o}$) at an altitude of 400 km, with a 31 day repeat, Fig. \ref{time_gravity} shows the CAI GG has to improve its sensitivity to several orders of magnitude in order to become sensitive to time-variable signals. Unless improving the sensitivity of the CAI (with longer arms or longer interferometer time) at the cost of a much bigger satellite and/or smaller measurement bandwidth, a GRACE-type mission should be more relevant for this type of missions. Nevertheless CAI could be useful for such a mission to calibrate electrostatic accelerometers as shown in the following section. 

\section{Hybridization concept for calibrating electrostatic accelerometers}

Future mission scenarios are already proposed for improving the performances provided by GOCE and GRACE. For instance NGGM from ESA and GRACE2 from NASA are two space mission concepts having the objective of monitoring the temporal variation of the Earth's gravity field at high resolution (similar to the spatial resolution provided by GOCE) by means of a laser interferometer ranging system in a Low-Low SST configuration \cite{Massotti}. Gradiometers or accelerometers with performances similar to those embarked on GOCE are essential for the success of such missions.

Inertial sensing technologies relying on electrostatic forces or on Atom Interferometer are identified as very good candidates for future space missions dedicated to Earth observation. Each of these two types of instruments have their own assets, which are, for electrostatic accelerometers (EA), their demonstrated short term sensitivity and their maturity regarding space environment. For Atom Interferometers the assets are, amongst others, the absolute nature of the measurement, a white noise spectrum even in the low frequency range and therefore no need for calibration processes. These two technologies seem in some aspects very complementary, and a hybrid sensor bringing together all their assets could be the opportunity to take a big step forward in this context. This could be achieved as it is realized in frequency measurements where quartz oscillators are phase locked on atomic or optical clocks \cite{Vanier}. This technique could correct the spectrally colored noise of the electrostatic accelerometers in the lower frequency range.

\begin{figure}
\centering \includegraphics[width=1\linewidth]{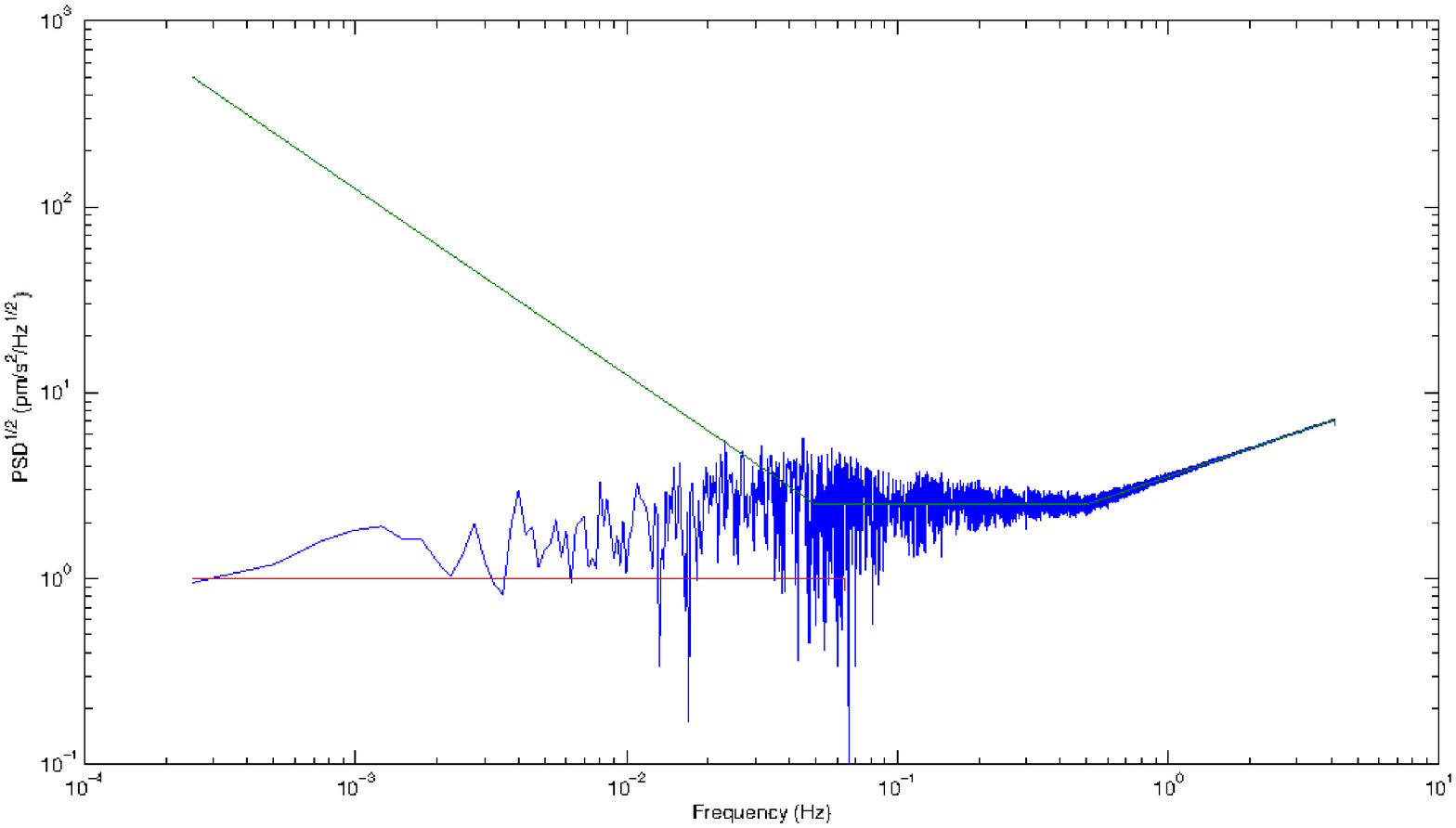}
\caption{Hybridization technique. In green PSD of an electrostatic accelerometer, in red PSD of a CAI accelerometer, in blue PSD of the hybrid signal.\label{hybrid}}
\end{figure}

To realize this hybridization, the mirror used for Raman transitions should be placed on the proof mass. Thus signal acquired by both instruments should be the same. As the frequency measurement is not the same for the EA (1 Hz) and for the CAI (0.1 Hz) the sensitivity function, which is very well characterized \cite{{syrte},{nagornyi}}, has to be applied on the EA signal. Then both signals can be compared and retroactively calibrate the EA for the low frequencies and still get the performances of the EA for the short term measurements (Fig. \ref{hybrid}).
Experimental hybridization has been recently done for gravity measurement on ground \cite{hybridSyrte} and shows this spectrally white behavior for a large bandwidth.

\section{Conclusion}
A concept of gravity gradiometer based on cold atom interferometer techniques is proposed. This instrument allows reaching sensitivity of $4.7\;mE/\sqrt{Hz}$, with the promise of a flat noise power spectral density also at low frequency, and a very high accuracy on rotation rates.

First estimations of an Earth gravity field model retrieved via the new gravity gradiometer concept have been evaluated taking into account different system parameters such as attitude control, altitude of the satellite, time duration of the mission, etc. It has been shown that if such an instrument could improve static gravity field measurements, a large effort has to be done to reach a significant improvement of the time variable ones. New techniques allow us to go beyond the Standard Quantum Limit (SQL) with squeezed state or Information-recycling beam splitters \cite{{Gross},{Haine}}. At best it is possible to reach the Heisenberg limit, i.e. to have a Signal-to-Noise Ratio (SNR) proportional to N. The phase noise could be extended at best to $1\;\mu rad$ which means up to 3 orders of magnitude improvement on the sensitivity of the CAI GG, on condition that other noise sources can be consistent with this limit.

In the meantime as long as this gravity gradiometer concept is not yet available, hybridization between quantum and classical techniques could be a promising option to improve the performance of accelerometers on next generation gravity missions. This technique could correct the spectrally colored noise of the electrostatic accelerometers in the low frequency range, as it has already been proven for ground gravity measurements.

\end{document}